\documentclass[crop]{CSL}
\usepackage{xcolor}
\usepackage{adjustbox}
\usepackage[version=4]{mhchem}

\jname{International Journal of Astrobiology}%

\begin{document}

\supertitle{Research Paper}

\title[Methylenimine towards G10.47+0.03]{Detection and prebiotic chemistry of possible glycine precursor molecule methylenimine towards the hot molecular core G10.47+0.03}

\author[Manna \& Pal]{Arijit Manna$^{1}$ and Sabyasachi Pal$^{1}$}

\address{\add{1}{Department of Physics and Astronomy, Midnapore City College, Paschim Medinipur, India 721129}}

\corres{\name{Sabyasachi Pal} \email{sabya.pal@gmail.com}}

\begin{abstract}
Amino acids are essential for the synthesis of protein. Amino acids contain both amine (R--NH$_{2}$) and carboxylic acid (R--COOH) functional groups, which help to understand the possible formation mechanism of life in the universe. Among the 20 types of amino acids, glycine (NH$_{2}$CH$_{2}$COOH) is known as the simplest non-essential amino acid. In the last 40 years, all surveys of NH$_{2}$CH$_{2}$COOH in the interstellar medium, especially in the star-formation regions, have failed at the millimeter and sub-millimeter wavelengths. We aimed to identify the possible precursors of NH$_{2}$CH$_{2}$COOH, because it is highly challenging to identify NH$_{2}$CH$_{2}$COOH in the interstellar medium. Many laboratory experiments have suggested that methylenimine (CH$_{2}$NH) plays a key role as a possible precursor of NH$_{2}$CH$_{2}$COOH in the star-formation regions via the Strecker synthesis reaction. After spectral analysis using the local thermodynamic equilibrium (LTE) model, we successfully identified the rotational emission lines of CH$_{2}$NH towards the hot molecular core G10.47+0.03 using the Atacama Compact Array (ACA). The estimated column density of CH$_{2}$NH towards G10.47+0.03 is (3.40$\pm$0.2)$\times$10$^{15}$ cm$^{-2}$ with a rotational temperature of 218.70$\pm$20 K, which is estimated from the rotational diagram. The fractional abundance of CH$_{2}$NH with respect to H$_{2}$ towards G10.47+0.03 is 2.61$\times$10$^{-8}$. We found that the derived abundance of CH$_{2}$NH agree fairly well with the existing two-phase warm-up chemical modelling abundance value of CH$_{2}$NH. We discuss the possible formation pathways of CH$_{2}$NH within the context of hot molecular cores, and we find that CH$_{2}$NH is likely mainly formed via neutral-neutral gas-phase reactions of CH$_{3}$ and NH radicals towards G10.47+0.03.\\\\
{\bf keywords}: {astrochemistry, Interstellar medium, prebiotic chemistry, star-formation region, millimeter astronomy}
\end{abstract}

\selfcitation{Manna, A \& Pal, S (2024). Detection of possible glycine precursor molecule methylenimine towards the hot molecular core G10.47+0.03. International Journal of Astrobiology}

\received{xx xxxx xxxx}

\revised{xx xxxx xxxx}

\accepted{xx xxxx xxxx}

\maketitle


\section{1. Introduction}
\label{sec:intro} 

\begin{figure*}
	\centering
	\includegraphics[width=0.9\textwidth]{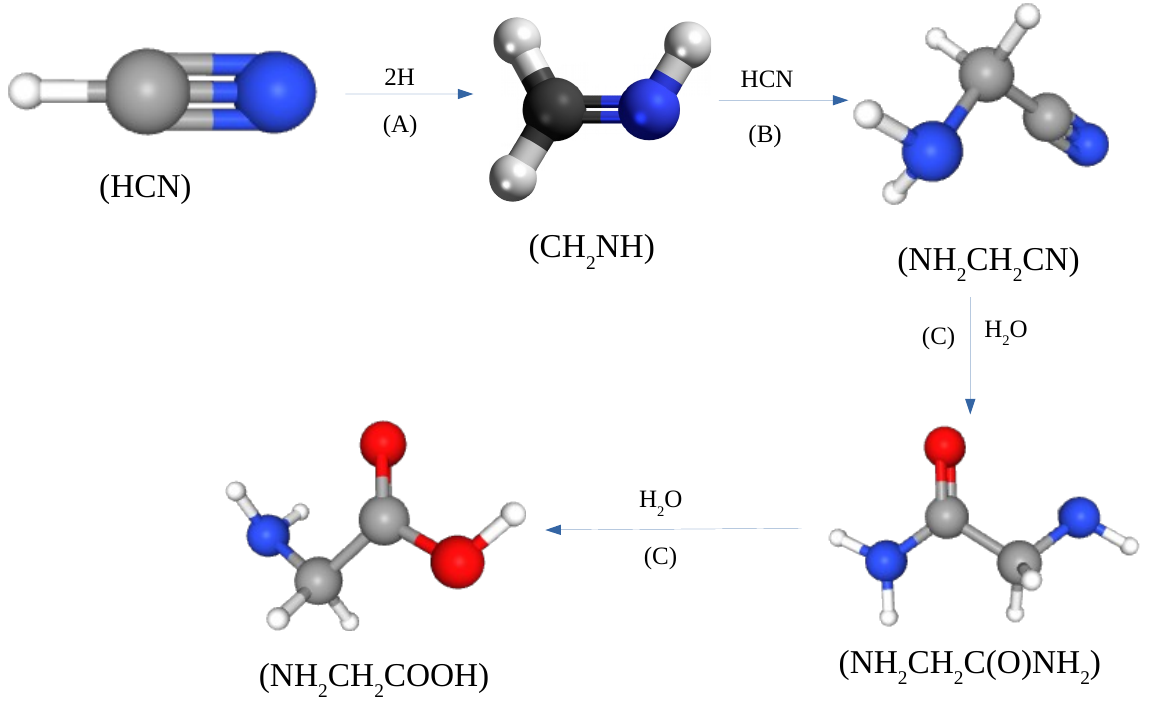}
	\caption{Proposed possible formation mechanism of CH$_{2}$NH and NH$_{2}$CH$_{2}$COOH. In the chemical diagram, the black/grey dumbbell indicates the carbon (C) atom, the white dumbbell indicates the hydrogen (H) atom, the blue dumbbell indicates the nitrogen (N) atom, and the red dumbbell indicates the oxygen (O) atom. In the chemical reaction, ``H$_{2}$O'' represents the hydrolysis process. References: (A) \citet{wo02}; \citet{thu11}; (B) \citet{dan11}; (C) \citet{alo18}.}
	\label{fig:reaction}
\end{figure*}

At millimeter and submillimeter wavelengths, approximately 290 prebiotic and complex organic molecules have been discovered in the interstellar medium (ISM) or circumstellar shells\footnote{\url{https://cdms.astro.uni-koeln.de/classic/molecules}}. The identification of complex prebiotic molecules in the ISM is important to understand the chemical evolution of biologically relevant prebiotic molecules from fundamental molecular species \citep{her09}. Hot molecular cores are one of the early stages of high-mass star-formation regions \citep{van98, her09, shi21, man23, man24a}. The early stages of the high-mass star-formation regions are known as the chemically rich phase, which plays an essential role in understanding the formation of chemical complexity in the ISM \citep{tan14, shi21}. In the hot molecular cores, the complex organic molecules escape from the icy surfaces of dust grains, or the complex organic molecules are created in the hot circumstellar gas \citep{her09}. Hot molecular cores are identified by their high gas density ($>$10$^{6}$ cm$^{-3}$), small source size ($<$0.1 pc), and warm temperature ($>$100 K) \citep{van98, kur00}. The warm-up time scale for hot molecular cores ranges from $\sim$10$^{4}$ to $\sim$10$^{6}$ years \citep{van98, gar06, gar13}. The hot molecular phase is characterized by the rich molecular spectra of several complex organic molecules such as methanol (CH$_{3}$OH) and methyl cyanide (CH$_{3}$CN) \citep{al17}. These complex molecules can form on the surface of dust grains on a cooler surface and are then released when the grains are heated owing to the formation of stars \citep{al17, man23}. Alternatively, these complex molecules may be created in massive young objects when the high temperature ($>$100 K) allows for endothermic reactions \citep{al17}. Therefore, both formation pathways of complex organic molecules are important for acquiring molecular abundance around the hot molecular cores. Higher spectral and spatial resolution observations are required to identify the different complex organic molecules and the spatial distribution of these molecules in hot molecular cores. The detection of disc candidates in hot molecular cores is extremely rare, implying a link between the hot molecular core chemistry and discs \citep{al17}. Studying the chemistry of the hot molecular cores of disc candidates can help us to understand the chemical evolution of high-mass star formations on small physical scales \citep{al17}.

The disc-like hot molecular core G10.47+0.03 is known as the ultra-compact (UC) H II region, which is located at a distance of 8.6 kpc with a luminosity of 5$\times$10$^{5}$ L$_{\odot}$ \citep{cer10, san14}. G10.47+0.03 is a disc-like candidate because, in this source, the hot core is embedded in the disc \citep{san14, man22a}. Earlier, \citet{rof11} conducted a molecular spectral line survey of G10.47+0.03, using the Submillimeter Array (SMA) telescope in the frequency range of 199.9--692.2 GHz. Using the LTE modelling, \citet{rof11} detected the rotational emission lines of several simple and complex organic molecules such as sulfur monoxide (SO), sulfur dioxide (SO$_{2}$), cyanide (CN), hydrogen cyanide (HCN), hydrogen isocyanide (HNC), formamide (NH$_{2}$CHO), cyanoacetylene (HC$_{3}$N), vinyl cyanide (C$_{2}$H$_{3}$CN), formaldehyde (H$_{2}$CO), ethynol (H$_{2}$C$_{2}$O), ethanol (C$_{2}$H$_{5}$OH), dimethyl ether (CH$_{3}$OCH$_{3}$), methyl formate (CH$_{3}$OCHO), methanol (CH$_{3}$OH), and acetone (CH$_{3}$COH$_{3}$) towards the G10.47+0.03. The rotational emission lines of methylamine (CH$_{3}$NH$_{2}$) and amino acetonitrile (NH$_{2}$CH$_{2}$CN) are also detected towards G10.47+0.03 \citep{oh19, man22a}. The CH$_{3}$NH$_{2}$ and NH$_{2}$CH$_{2}$CN molecules are known to be other possible precursors of the simplest amino acid, NH$_{2}$CH$_{2}$COOH, towards hot molecular cores. The emission lines of cyanamide (NH$_{2}$CN) and ethyl cyanide (C$_{2}$H$_{5}$CN) are detected from the hot molecular core G10.47+0.03 using ALMA \citep{man22b, man23a}. Recently, the rotational emission line of phosphorus nitride (PN) is detected towards the G10.47+0.03 \citep{man24b}.

The asymmetric top-molecule methylenimine (CH$_{2}$NH) is known to be a possible precursor of NH$_{2}$CH$_{2}$COOH in the ISM. The CH$_{2}$NH molecule was created by the hydrogenation of HCN on the dust surface of hot molecular cores \citep{wo02, thu11}. When CH$_{2}$NH and HCN both react with each other via the Strecker synthesis reaction, the complex amino and nitrile-bearing molecule amino acetonitrile (NH$_{2}$CH$_{2}$CN) is produced \citep{dan11}. The hydrolysis of NH$_{2}$CH$_{2}$CN on the grain surface created glycinamide (NH$_{2}$CH$_{2}$C(O)NH$_{2}$) \citep{alo18}. Finally, NH$_{2}$CH$_{2}$COOH can be created via hydrolysis of NH$_{2}$CH$_{2}$C(O)NH$_{2}$ on the grain surface of hot molecular cores \citep{alo18}. The proposed formation processes for CH$_{2}$NH and NH$_{2}$CH$_{2}$COOH are shown in Figure ~\ref{fig:reaction}. Earlier, \citet{suz16} claimed that the CH$_{2}$NH molecule is created in the gas phase between the reactions of CH$_{3}$ and NH (CH$_{3}$ + NH $\rightarrow$ CH$_{2}$NH). Quantum chemical studies have shown that \ce{CH3NH2} is produced via the sequential hydrogenation of \ce{CH2NH} \citep{jo22}. This indicates that both \ce{CH2NH} and \ce{CH3NH2} are chemically linked in ISM \citep{jo22}. Subsequently, \citet{gar22} confirmed that both \ce{CH2NH} and \ce{CH3NH2} are chemically linked towards hot molecular cores by using three-phase warm-up chemical models. Both \citet{jo22} and \citet{gar22} showed that \ce{CH3NH2} and \ce{CH2NH} are chemically connected with \ce{NH2CH2COOH}. Previously, many authors claimed that CH$_{2}$NH was detected in the high-mass star-formation region Sgr B2. Evidence of CH$_{2}$NH was found near Sgr B2 (OH) \citep{god73, tur89}, Sgr B2 (N) \citep{hal13}, and Sgr B2 (M) \citep{su91}. Previously, \citet{jo08, jo11} created the spatial distribution of CH$_{2}$NH from Sgr B2 (N) to Sgr B2 (S) at wavelengths of 3 and 7 mm using the MOPRA telescope. \citet{bel13} demonstrated a detailed analysis of CH$_{2}$NH towards Sgr B2 (N) and Sgr B2 (M) using the IRAM 30 m telescope. The rotational emission lines of CH$_{2}$NH were also detected for W51 e1/e2, Orion KL, G34.3+0.15, G19.61-0.23, IRAS 16293--2422 B, and NGC 6334I \citep{dic97, whi03, qin10, lig18, bo19}. Recently, CH$_{2}$NH megamaser\footnote{A megamaser is a type of astrophysical maser in which the luminosities of the spectral lines are 100 million times brighter than normal masers emission lines in the ISM.} lines were detected in six compact obscured nuclei using the Very Large Array (VLA) \citep{gor21}.

In this article, we present the identification of the possible NH$_{2}$CH$_{2}$COOH precursor molecule CH$_{2}$NH towards G10.47+0.03, using ACA. To estimate the column density and rotational temperature of CH$_{2}$NH, we used a rotational diagram model. ACA observations and data reductions are presented in Section~2. The results of the detection of the emission lines of CH$_{2}$NH are presented in Section~3. The discussion and conclusion of the detection of CH$_{2}$NH are presented in Sections ~4 and 5.

\section{2. Observation and data reductions}
\label{obs}
We used the archival data of G10.47+0.03 in cycle 4, which was observed using the Atacama Compact Array (ACA) with a 7-m array (PI: Rivilla, Victor; ID: 2016.2.00005.S). The ACA is the heart of the Atacama Large Millimeter/submillimeter Array (ALMA). The observed phase centre of the hot molecular core G10.47+0.03 is ($\alpha,\delta$)$_{\rm J2000}$ = 18:08:38.232, --19:51:50.400. The observation was carried out on September 16, 2017, using 11 antennas. The observations were made with ACA band 4 with spectral ranges of 127.47--128.47 GHz, 129.74--130.74 GHz, 139.07--140.07 GHz, and 140.44--141.44 GHz and a corresponding spectral resolution of 488 kHz. During the observation, the flux calibrator and bandpass calibrator were J1924--2914, and the phase calibrator was J1833--210B.
	
For data reduction and imaging, we used the Common Astronomy Software Application (CASA 5.4.1) with an ALMA data reduction pipeline \citep{mc07}. The data analysis flow chart is shown in \citet{man24c}. For flux calibration using the flux calibrator, we used the Perley-Butler 2017 flux calibrator model for each baseline to scale the continuum flux density of the flux calibrator using the CASA task {SETJY} \citep{per17}. We constructed the flux and bandpass calibration after flagging bad antenna data and channels using the CASA pipeline with tasks {hifa\_bandpassflag} and {hifa\_flagdata}. After the initial data reduction, we used the task {MSTRANSFORM} with all available rest frequencies to separate the target data of G10.47+0.03. For continuum and background subtraction, we used task {UVCONTSUB} in the UV plane of the separated calibrated data. We used the CASA task {TCLEAN} with a {Briggs} weighting robust value of 0.5, to create continuum and spectral images of the G10.47+0.03. To produce spectral images, we used the {SPECMODE = CUBE} parameter in the TCLEAN task. The final spatial resolution of the spectral data cubes was 10.48$^{\prime\prime}\times$6.28$^{\prime\prime}$, 10.82$^{\prime\prime}\times$6.39$^{\prime\prime}$, 12.08$^{\prime\prime}\times$6.90$^{\prime\prime}$, and 12.08$^{\prime\prime}\times$6.79$^{\prime\prime}$ between the frequency ranges of 127.47--128.47 GHz, 129.74--130.74 GHz, 139.07--140.07 GHz, and 140.44--141.44 GHz with a spectral resolution of 488.28 kHz. Finally, we used the CASA task {IMPBCOR} to correct the primary beam pattern in continuum images and spectral data cubes.

\begin{figure*}
	\centering
	\includegraphics[width=1.0\textwidth]{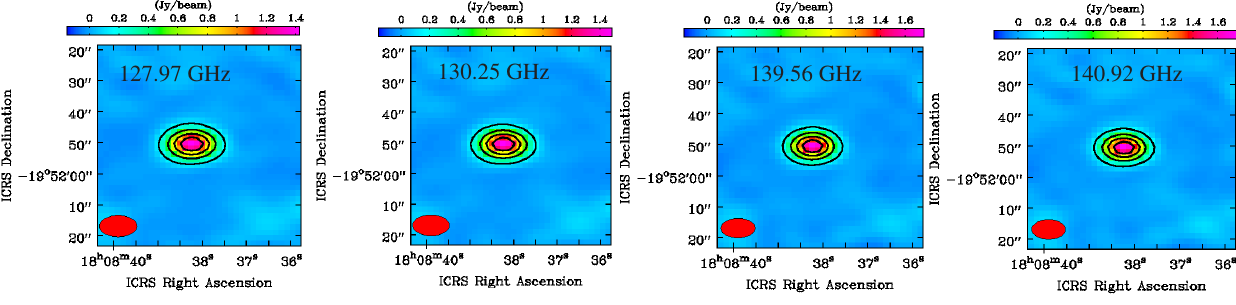}
	\caption{Millimeter-wavelength continuum emission images of the hot molecular core G10.47+0.03. Continuum emission images are obtained with ACA band 4 at frequencies of 127.97 GHz, 130.25 GHz, 139.56 GHz, and 140.92 GHz. The contour levels start at 3$\sigma$, where $\sigma$ is the RMS of each continuum image. The contour levels increase by a factor of $\surd$2. The red circles indicate the synthesized beams of the continuum images. The corresponding synthesized beam sizes and RMS values of all continuum images are presented in Table~\ref{tab:cont}.}
	\label{fig:continuum}
\end{figure*}

\section{3. Result}
\label{res}
\subsection{3.1 Continuum emission towards the G10.47+0.03}
We presented the continuum emission images of the hot molecular core G10.47+0.03 at frequencies of 127.97 GHz, 130.25 GHz, 139.56 GHz, and 140.92 GHz. The continuum images are shown in Figure~\ref{fig:continuum}, where the surface brightness colour scale has units of the Jy beam$^{-1}$. After the creation of the continuum emission images, we fitted the 2D Gaussian over the continuum emission images using the CASA task {IMFIT} and estimated the integrated flux density in Jy, peak flux density in Jy beam$^{-1}$, synthesized beam size in arcsec ($^{\prime\prime}$), deconvolved beam size in arcsec ($^{\prime\prime}$), position angle in degrees ($^{\circ}$), and RMS in mJy of the hot core G10.47+0.03. The estimated continuum image properties of hot core G10.47+0.03 are shown in Table~\ref{tab:cont}. We noticed that the continuum emission region of G10.47+0.03 is smaller than the synthesized beam size, which was estimated after fitting the 2D Gaussian over the continuum emission region. This indicates that the continuum emission image of G10.47+0.03 was not resolved between the frequency range of 127.97 GHz to 140.92 GHz. Recently, \citet{man23a} reported the detection of continuum emission from the G10.47+0.03 in the frequency range of 130.23 GHz--160.15 GHz with a flux density variation of 1.36--2.71 Jy.

\begin{table*}
	\centering
	\caption{Summary of the millimeter wavelength continuum images of G10.47+0.03.}
	\begin{adjustbox}{width=1.0\textwidth}
		\begin{tabular}{|c|c|c|c|c|c|c|c|c|c|c|c|c|c|c|c|c|}
			\hline 
			Frequency&Wavelength &Integrated flux & Peak flux &Beam size&Deconvolved source size &RMS & Position angle\\
			(GHz) &(mm)     & (Jy)          &  (Jy beam$^{-1}$) &($^{\prime\prime}$$\times$$^{\prime\prime}$)&($^{\prime\prime}$$\times$$^{\prime\prime}$) &(mJy)&($^{\circ}$)\\
			\hline
			127.97&2.34&1.57$\pm$0.01&1.44$\pm$0.08&12.02$\times$6.87 &2.60$\times$2.18&9.16&--89.71\\
			130.25&2.30&1.66$\pm$0.01&1.51$\pm$0.08&11.96$\times$6.73 &2.73$\times$2.52&8.76&--89.63\\		
			139.56&2.14&1.98$\pm$0.02&1.78$\pm$0.01&10.73$\times$6.35 &2.71$\times$2.29&10.72&--89.61\\		
			140.92 &2.12&2.00$\pm$0.02&1.78$\pm$0.01&10.61$\times$6.25&2.75$\times$2.44&11.68&--89.91\\
			\hline 
		\end{tabular}	
	\end{adjustbox}
	\label{tab:cont}
\end{table*}

\subsection{3.2 Identification of the CH$_{2}$NH in the G10.47+0.03}
\label{sec:fitting}
\begin{figure*}
	\centering
	\includegraphics[width=1.0\textwidth]{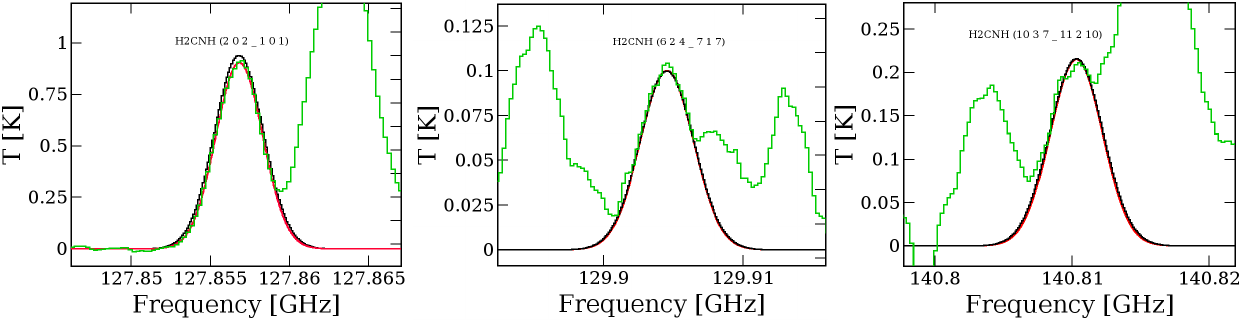}
	\caption{Identified rotational emission lines of CH$_{2}$NH towards the G10.47+0.03 in the frequency ranges of 127.47--128.47 GHz, 129.74--130.74 GHz, and 140.44--141.44 GHz. The green spectra indicate the millimeter-wavelength molecular spectra of G10.47+0.03. The black spectra present the best-fit LTE model spectra of CH$_{2}$NH, and the red spectra indicate the Gaussian model. The radial velocity of the spectra is 68.50 km s$^{-1}$.}
	\label{fig:ltespec}
\end{figure*}

First, we extracted the millimeter-wavelength molecular spectra from the spectral data cubes to create a 23.75$^{\prime\prime}$ diameter circular region over the G10.47+0.03. The synthesized beam sizes of the spectral data cubes of hot core G10.47+0.03 is 10.48$^{\prime\prime}\times$6.28$^{\prime\prime}$, 10.82$^{\prime\prime}\times$6.39$^{\prime\prime}$, 12.08$^{\prime\prime}\times$6.90$^{\prime\prime}$, and 12.08$^{\prime\prime}\times$6.79$^{\prime\prime}$. Hot core G10.47+0.03 is located at a distance of 8.6 kpc and at that distance, a $\sim$10$^{\prime\prime}$ resolution refers to a spatial scale of 0.4 pc. This implies that the extracted spectrum mostly represents the outer envelope. The systematic velocity ($V_{LSR}$) of G10.47+0.03 is 68.50 km s$^{-1}$ \citep{rof11}. We used the second-order polynomial to subtract the baseline of the entire spectra. To identify the rotational emission lines of CH$_{2}$NH, we used the local thermodynamic equilibrium (LTE) model with the Cologne Database for Molecular Spectroscopy (CDMS) database \citep{mu05}. For LTE modelling, we used CASSIS \citep{vas15}. The LTE assumptions are valid in the inner region of G10.47+0.03 because the gas density of the warm inner region of the hot core was 7$\times$10$^{7}$ cm$^{-3}$ \citep{rof11}. To fit the LTE model spectra of CH$_{2}$NH over the millimeter wavelength spectra of G10.47+0.03, we applied the Markov Chain Monte Carlo (MCMC) algorithm in CASSIS. After the LTE analysis, we have detected a total of three transitions of CH$_{2}$NH i.e., $J$ = 2(0,2)--1(0,1), $J$ = 6(2,4)--7(1,7), and $J$ = 10(3,7)--11(2,10). The three detected transitions of CH$_{2}$NH had hyperfine lines. We do not discuss the hyperfine lines regarding the identified transitions of CH$_{2}$NH because the current spectral resolution is insufficient to resolve the hyperfine lines. The CH$_{2}$NH is a simple asymmetric top with all atoms being on the simple plane, and the transitions are described using labels of $J^{\prime}$, $K_{p}^{\prime}$, $K_{o}^{\prime}$, and $J^{\prime\prime}$, $K_{p}^{\prime\prime}$, $K_{o}^{\prime\prime}$. In the transition of CH$_{2}$NH, $J$ indicated the total rotational angular momentum quantum number, $K_{p}$ indicated the projection of $J$ on the symmetry axis in the limiting prolate symmetric top, $K_{o}$ indicated the projection of $J$ on the symmetry axis in the limiting oblate symmetric top, and $F$ indicated the total angular momentum quantum number, which includes the nuclear spin for the nucleus with the largest $\chi$ or eQq where $\chi$ or eQq denoted the nuclear electric quadrupole coupling constant along the indicated principal axis \citep{kir73}. There were no missing transitions of CH$_{2}$NH in the observable frequency ranges. As per the CDMS and online molecular database Splatalogue, we find that all the detected transitions of \ce{CH2NH} are not blended with other nearby molecular transitions. Using the LTE model, the best-fit column density of CH$_{2}$NH was (3.21$\pm$1.5)$\times$10$^{15}$ cm$^{-2}$ with an excitation temperature of 210.50$\pm$32.82 K and a source size of 10.78$^{\prime\prime}$. The full-width half maximum (FWHM) of the LTE-fitted rotational emission spectra of CH$_{2}$NH was 9.5 km s$^{-1}$. The LTE-fitted rotational emission spectra of CH$_{2}$NH are shown in Figure ~\ref{fig:ltespec}. After identifying the rotational emission lines of CH$_{2}$NH using the LTE model, we obtained the molecular transitions, upper-state energy ($E_u$) in K, Einstein coefficients ($A_{ij}$) in s$^{-1}$, line intensity ($S\mu^{2}$) in Debye$^{2}$, and optical depth ($\tau$). We also verified the detected transitions of CH$_{2}$NH from \citet{kir73}. To estimate the proper FWHM and integrated intensity ($\rm{\int T_{mb}dV}$) of the detected emission lines of CH$_{2}$NH, we fitted a Gaussian model to the observed spectra of CH$_{2}$NH. A summary of the detected transitions and spectral line properties of CH$_{2}$NH are presented in Table~\ref{tab:MOLECULAR DATA}. 

\begin{table*}
	\centering
	\scriptsize 
	\caption{Summary of the molecular line parameters of the CH$_{2}$NH towards the G10.47+0.03}
	\begin{adjustbox}{width=1.0\textwidth}
		\begin{tabular}{|c|c|c|c|c|c|c|c|c|c|c|c|c|c|c|c|c|}
			\hline 
			Frequency &Transition$^{*}$& $E_{u}$$^{\Delta}$ & $A_{ij}$$^{\Delta}$ &g$_{up}$$^{\Delta}$&$S\mu^{2}$$^{\dagger}$&FWHM$^{\ddag}$&$\rm{\int T_{mb}dV}$$^{\ddag}$ &Optical depth\\
			
			(GHz) &$J^{\prime}$($K_{p}^{\prime}$,$K_{o}^{\prime}$)--$J^{\prime\prime}$($K_{p}^{\prime\prime}$,$K_{o}^{\prime\prime}$)&(K)&(s$^{-1}$) & &(Debye$^{2}$)&(km s$^{-1}$)&(K km s$^{-1}$)  &($\tau$) \\
			\hline
			127.856&2(0,2)--1(0,1)&9.21&1.75$\times$10$^{-5}$&15&10.76&9.49$\pm$0.10 &5.97$\pm$0.25 &2.96$\times$10$^{-4}$\\
			129.904&6(2,4)--7(1,7)&96.38&3.23$\times$10$^{-6}$&39&4.93 &9.62$\pm$0.48&0.61$\pm$0.06&4.87$\times$10$^{-4}$\\
			140.810&10(3,7)--11(2,10)&240.27&4.85$\times$10$^{-6}$&63&9.41 &9.56$\pm$0.98 &0.23$\pm$0.09 &8.51$\times$10$^{-4}$\\
			
			\hline
		\end{tabular}	
	\end{adjustbox}
	\label{tab:MOLECULAR DATA}\\
	{{*}}--All detected transitions of CH$_{2}$NH are also verified from Table~2 in \cite{kir73} and CDMS molecular database.\\
	
    $\Delta$--The values of E$_{up}$, A$_{ij}$ and g$_{up}$ are taken from the CDMS database.\\
	$\dagger$--The line intensity $S\mu^{2}$ is defined by the product of the transition line strength $S$ and square of the dipole moment $\mu^{2}$ of CH$_{2}$NH. The values of $S\mu^{2}$ of detected transitions of CH$_{2}$NH are taken from online molecular database \href{https://splatalogue.online/index.php}{splatalogue}. \\
	${\ddag}$--FWHM and $\rm{\int T_{mb}dV}$ are estimated from the fitting of the Gaussian model over the observed spectra of CH$_{2}$NH\\.
\end{table*}

\subsection{3.3 Rotational diagram analysis of CH$_{2}$NH}
\label{sec:rotd}
\begin{figure}
	\centering
	\includegraphics[width=0.48\textwidth]{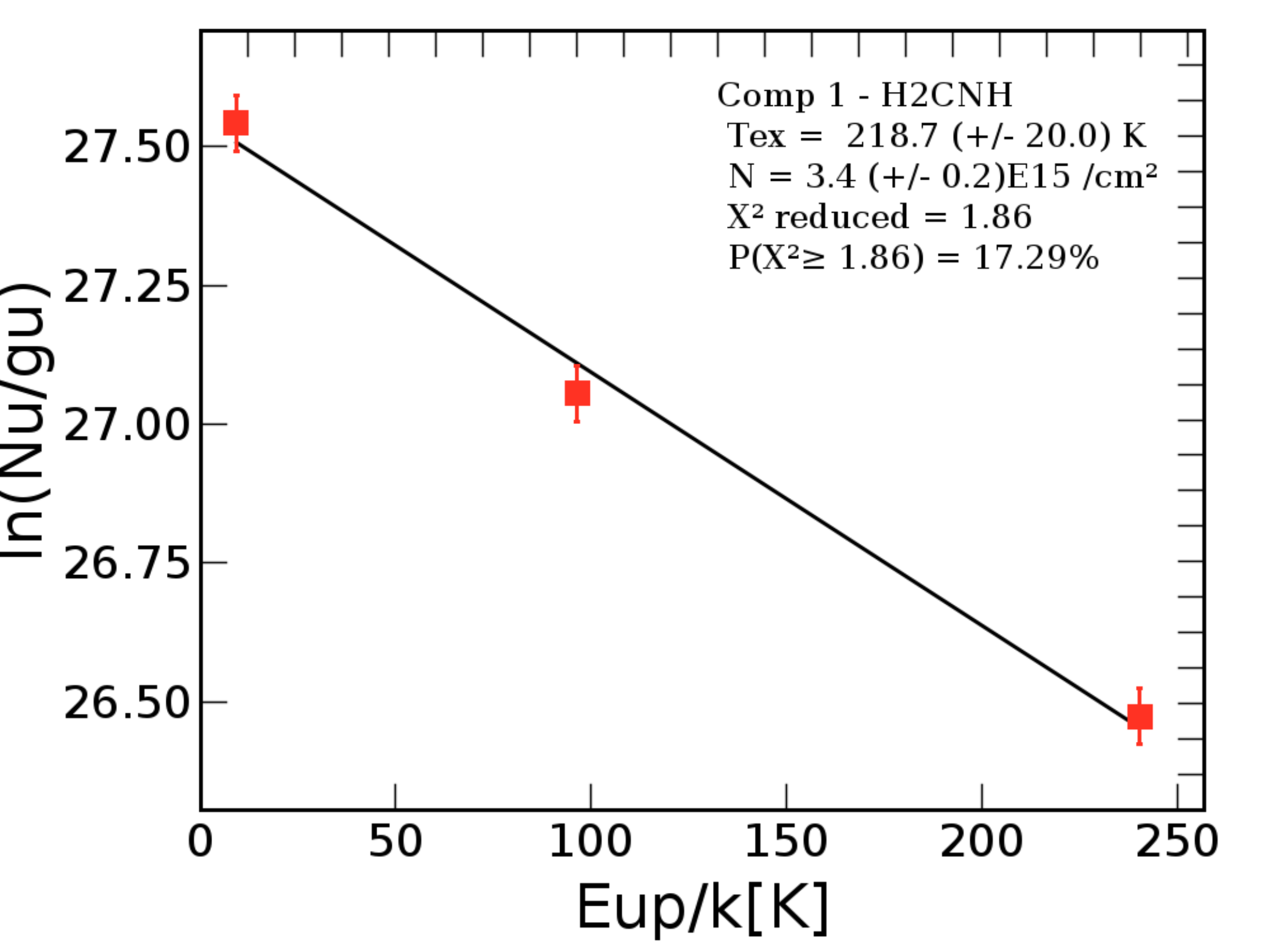}
	\caption{Rotational diagram of CH$_{2}$NH towards G10.47+0.03. In the rotational diagram, the red blocks represent the statistical data points of all detected transitions, and the solid black line indicates the fitted straight line, which helps estimate the column density and rotational temperature of CH$_{2}$NH.}
	\label{fig:rotd}
\end{figure}

In this work, we used the rotational diagram method to estimate the total column density ($N$) in cm$^{-2}$ and the rotational temperature ($T_{rot}$) in K of CH$_{2}$NH because we detected multiple transition lines of CH$_{2}$NH towards G10.47+0.03. Initially, we assumed that the detected CH$_{2}$NH emission lines were optically thin and populated under the LTE conditions. The equation of column density for optically thin molecular emission lines can be expressed as \citep{gol99},
\begin{equation}
{N_u^{thin}}=\frac{3{g_u}k_B\int{T_{mb}dV}}{8\pi^{3}\nu S\mu^{2}}
\end{equation}
where $g_u$ is the degeneracy of the upper state, $\mu$ is the electric dipole moment, $S$ indicates the strength of the transition lines, $\nu$ is the rest frequency, $k_B$ is Boltzmann's constant, and $\int T_{mb}dV$ indicates the integrated intensity. The total column density of CH$_{2}$NH under LTE conditions can be written as,
\begin{equation}
\frac{N_u^{thin}}{g_u} = \frac{N_{total}}{Q(T_{rot})}\exp(-E_u/k_BT_{rot})
\end{equation}
where $E_u$ is the upper-state energy of CH$_{2}$NH, $T_{rot}$ is the rotational temperature of CH$_{2}$NH, and ${Q(T_{rot})}$ is the partition function at the extracted rotational temperature. The rotational partition function of CH$_{2}$NH at 75 K is 740.457, that at 150 K is 2084.970, and that at 300 K is 5892.504 \citep{mu05}. Equation 2 can be rearranged as,
\begin{equation}
ln\left(\frac{N_u^{thin}}{g_u}\right) = ln(N)-ln(Q)-\left(\frac{E_u}{k_BT_{rot}}\right)
\end{equation}
Equation 3 indicates a linear relationship between the upper state energy ($E_{u}$) and $\ln(N_{u}/g_{u})$ of CH$_{2}$NH. The value $\ln(N_{u}/g_{u})$ was estimated using Equation 1. Equation 3 indicates that the spectral parameters with respect to the different transition lines of CH$_{2}$NH should be fitted with a straight line, whose slope is inversely proportional to the rotational temperature ($T_{rot}$), with its intercept yielding $\ln(N/Q)$, which will help estimate the molecular column density of CH$_{2}$NH. For the rotational diagram analysis, we estimated the spectral line parameters of CH$_{2}$NH after fitting the Gaussian model over the observed spectra of CH$_{2}$NH using the Levenberg-Marquardt algorithm in CASSIS (see Section~3.2 for details on spectral fitting). For the rotational diagram analysis, we used all the detected transitions of CH$_{2}$NH to estimate the accurate column density and rotational temperature because all the detected transitions of CH$_{2}$NH are non-blended. The resultant rotational diagram of CH$_{2}$NH is shown in Figure~\ref{fig:rotd}, which was created using the {ROTATIONAL DIAGRAM} module in CASSIS. In the rotational diagram, the vertical red error bars indicate the absolute uncertainty of $\ln(N_{u}/g_{u})$, which was determined from the estimated error of $\int T_{mb}dV$. From the rotational diagram, we estimated the column density of CH$_{2}$NH to be (3.40$\pm$0.2)$\times$10$^{15}$ cm$^{-2}$ with a rotational temperature of 218.70$\pm$20 K. From the LTE spectral modelling, we found that the column density and excitation temperature of \ce{CH2NH} are (3.21$\pm$1.5)$\times$10$^{15}$ cm$^{-2}$ and 210.50$\pm$32.82 K, which are nearly similar to the estimated column density and rotational temperature of \ce{CH2NH} using the rotational diagram model. Our derived rotational temperature of CH$_{2}$NH indicates that the detected transitions of CH$_{2}$NH arise from the warm-inner region of G10.47+0.03 because the temperature of the hot molecular core is above $\geq$100 K \citep{van98}. To determine the fractional abundance of \ce{CH2NH}, we use the column density of \ce{CH2NH} inside the 12.08$^{\prime\prime}$ beam and divide it by the column density of \ce{H2}. The estimated fractional abundance of CH$_{2}$NH towards G10.47+0.03 with respect to H$_{2}$ was 2.61$\times$10$^{-8}$, where the column density of H$_{2}$ towards G10.47+0.03 was 1.30$\times$10$^{23}$ cm$^{-2}$ \citep{suz16}.

\subsection{3.4 Spatial distribution of CH$_{2}$NH in the G10.47+0.03}
\begin{figure*}
	\centering
	\includegraphics[width=1.0\textwidth]{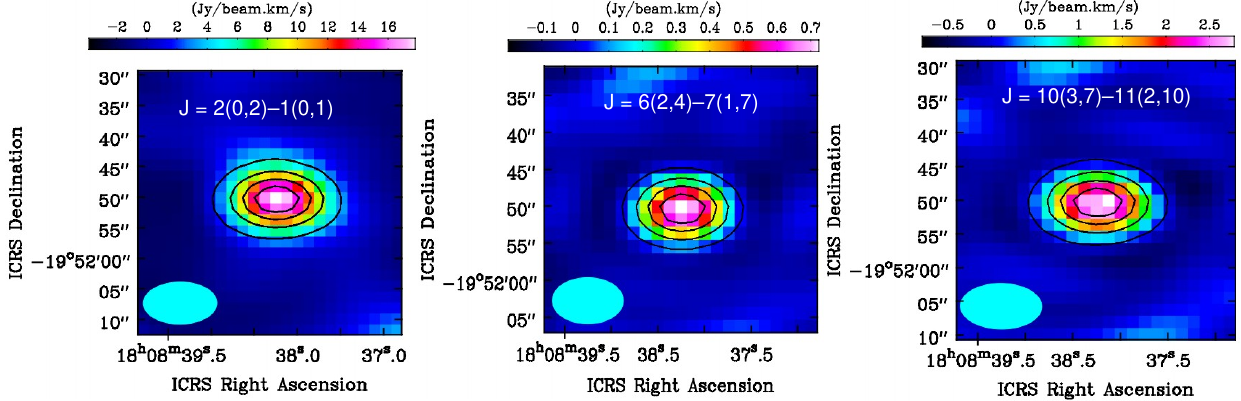}
	\caption{Integrated emission maps of detected transitions of CH$_{2}$NH towards the G10.47+0.03, which are overlaid with the 2.34 mm continuum emission map. The contour levels are 20\%, 40\%, 60\%, and 80\% of peak flux. The cyan circle represents the synthesized beam of the integrated emission maps.}
	\label{fig:emissionmap}
\end{figure*}

We created integrated emission maps of CH$_{2}$NH towards G10.47+0.03, using the CASA task {IMMOMENTS}. Integrated emission maps of CH$_{2}$NH were created by integrating the spectral data cubes in the velocity ranges of 61.06--74.11 km s$^{-1}$, 63.80--75.28 km s$^{-1}$, and 64.40--70.93 km s$^{-1}$, where the emission lines of CH$_{2}$NH were detected. We created integrated emission maps for the three non-blended transitions of CH$_{2}$NH towards G10.47+0.03. The integrated emission maps of CH$_{2}$NH with different transitions towards G10.47+0.03 are shown in Figure~\ref{fig:emissionmap}. The resultant integrated emission maps of CH$_{2}$NH were overlaid with the 2.34 mm continuum emission map of G10.47+0.03. The integrated emission maps of CH$_{2}$NH exhibit a peak at the continuum position. From the integrated emission maps, it is evident that the transitions of the CH$_{2}$NH molecule arise from the warm inner part of the hot core region of G10.47+0.03. This indicates that the temperature of the detected transition lines of \ce{CH2NH} is above 100 K because the temperature of the hot core $\geq$100 K \citep{van98}. From the rotational diagram, we estimate the temperature of \ce{CH2NH} is 218.7$\pm$20 K, which indicates the emission lines of \ce{CH2NH} arise from the warm-inner region of G10.47+0.03. After the generation of the integrated emission maps of all identified lines of CH$_{2}$NH, we estimated the emitting regions of CH$_{2}$NH towards G10.47+0.03 by fitting the 2D Gaussian over the integrated emission maps of CH$_{2}$NH using the CASA task {IMFIT}. The deconvolved beam size of the emitting region of CH$_{2}$NH was estimated by the following equation,

\begin{equation}
\theta_{S}=\sqrt{\theta^2_{50}-\theta^2_{beam}}
\end{equation}
where $\theta_{50} = 2\sqrt{A/\pi}$ indicates the diameter of the circle whose area ($A$) corresponds to the 50\% line peak of CH$_{2}$NH and $\theta_{beam}$ is the half-power width of the synthesized beam \citep{riv17, man22c, man23c}. The estimated emitting regions of the $J$ = 2(0,2)--1(0,1), $J$ = 6(2,4)--7(1,7), and $J$ = 10(3,7)--11(2,10) transitions of CH$_{2}$NH were 10.520$^{\prime\prime}$ (0.44 pc), 10.275$^{\prime\prime}$ (0.430 pc), and 10.781$^{\prime\prime}$ (0.451 pc). The emitting region of CH$_{2}$NH varied between 10.520$^{\prime\prime}$ and 10.781$^{\prime\prime}$. We observed that the emitting regions of CH$_{2}$NH were similar or small with respect to the synthesized beam size of the integrated emission maps, which means that the transition lines of CH$_{2}$NH were not spatially resolved or, at best, marginally resolved. Therefore, we cannot draw any conclusions about the morphology of the integrated emission maps of CH$_{2}$NH towards G10.47+0.03. Higher spectral resolution observations were needed using the ALMA 12 m array to solve the spatial distribution morphology of CH$_{2}$NH towards the hot molecular core G10.47+0.03.

\section{4. Discussion}
\label{dis}
\subsection{4.1. CH$_{2}$NH towards the G10.47+0.03}
We presented the first interferometric detection of possible \ce{NH2CH2COOH} precursor molecule CH$_{2}$NH towards the G10.47+0.03 using the ACA band 4. We identified a total of three transition lines of CH$_{2}$NH towards G10.47+0.03, and after spectral analysis using the LTE model, we observed that all identified transitions of CH$_{2}$NH are non-blended. Subsequently, these non-blended transition lines were used for rotational diagram analysis to estimate the total column density and rotational temperature of CH$_{2}$NH. Earlier, \citet{suz16} first attempted to search the rotational emission lines of CH$_{2}$NH from G10.47+0.03 and other hot molecular core objects. \citet{suz16} identified three transition lines of CH$_{2}$NH i.e., $J$ = 4(0,4)--3(1,3), $J$ = 4(1,4)--3(1,3), and $J$ = 4(2,3)--3(2,2) towards the G10.47+0.03 using the Nobeyama Radio Observatory (NRO) 45 m single dish telescope. Many questions arise in the detection of CH$_{2}$NH towards G10.47+0.03 by \citet{suz16}. First, all detected spectral lines of CH$_{2}$NH towards G10.47+0.03 were below 2.5$\sigma$ statistical significance (for details, see Figure 3 in \citet{suz16}), and the authors did not use any radiative transfer model for spectral characterization of CH$_{2}$NH. \citet{suz16} also did not discuss the blending effect of CH$_{2}$NH with nearby molecular transitions in the molecular spectra of G10.47+0.03. The single-dish observation of CH$_{2}$NH by \citet{suz16} could not study the spatial distribution of CH$_{2}$NH towards G10.47+0.03. Thus, \citet{suz16} did not estimate any information regarding the source size or emitting region of CH$_{2}$NH, which restricted the accuracy of their measurements of the proper column density of the detected molecules. We also observed that the upper state energy ($E_{u}$) of the transition lines of CH$_{2}$NH detected by \cite{suz16} varies between 10 and 30 K. In the rotational diagram, lower energy levels will not enable accurate determination of the column density. In the rotational diagram, transitions at significantly higher energy levels can determine a more accurate column density. Our interferometric detection of CH$_{2}$NH using ACA gives us confidence in the more accurate column density of CH$_{2}$NH because the upper state energies of the detected transitions vary between 9 K and 240 K. From the spatial distribution analysis, we estimated that the emission regions of CH$_{2}$NH vary between 10.520$^{\prime\prime}$ --10.781$^{\prime\prime}$. The estimated molecular column density of CH$_{2}$NH towards G10.47+0.03 using the ACA was (3.40$\pm$0.2)$\times$10$^{15}$ cm$^{-2}$ with a rotational temperature of 218.70$\pm$20 K. Earlier, \citet{suz16} estimated that the column density of CH$_{2}$NH towards G10.47+0.03 using the NRO telescope is (4.70$\pm$1.6)$\times$10$^{15}$ cm$^{-2}$ with a rotational temperature of 84$\pm$57 K. We find that the our estimated column density of CH$_{2}$NH is similar to that reported by \citet{suz16}, but the temperature is different. The estimated rotational temperature of \ce{CH2NH} by \citet{suz16} indicates that the detected transition lines of \ce{CH2NH} arise from the cold region of G10.47+0.03. Our estimated temperature indicates that the emission lines of \ce{CH2NH} arise from the warm-inner region of G10.47+0.03. \citet{suz16} found the lower rotational temperature due to the lower spatial and spectral resolution of the NRO telescope. Our estimated higher excitation temperature of \ce{CH2NH} using ACA is accurate because the temperature of the hot molecular cores is above 100 K \citep{van98}. The other two precursors of \ce{NH2CH2COOH}, such as \ce{CH3NH2} \citep{oh19} and \ce{NH2CH2CN} \citep{man22a} were also detected towards G10.47+0.03 using the NRO and ALMA telescopes. The detection of \ce{CH2NH} towards G10.47+0.03 indicates that three possible precursors of \ce{NH2CH2COOH} are present in G10.47+0.03. That means the hot molecular core G10.47+0.03 is an ideal candidate for searching the emission lines of \ce{NH2CH2COOH}.

\subsection{4.2. Comparison with modelled and observed abundance of CH$_{2}$NH}
After estimating the fractional abundance of CH$_{2}$NH towards G10.47+0.03, we compared the estimated abundance of CH$_{2}$NH with the modelled abundance of CH$_{2}$NH, which was estimated from the two-phase warm-up chemical model \citep{suz16}. For chemical modelling, \cite{suz16} used the gas-grain chemical kinetics code NAUTILUS in an environment with hot molecular cores. In chemical modelling, \citet{suz16} assumed an isothermal collapse phase after a static warm-up phase. In the first phase, the gas density rapidly increased from 3$\times$10$^{3}$ to 1$\times$10$^{7}$ cm$^{-3}$, and under free-fall collapse, the dust temperature decreased from 16 to 8 K. In the second phase, the gas density remained constant at 1$\times$10$^{7}$ cm$^{-3}$ and the gas temperature fluctuated rapidly from 8 K to 400 K \citep{suz16}. In chemical modelling, \citet{suz16} used the neutral-neutral reaction between \ce{CH3} and NH radicals in the gas phase and the neutral-neutral reaction between \ce{CH2} and NH radicals on the grain surface to create CH$_{2}$NH under the condition of hot molecular cores. The gas temperature of G10.47+0.03 was $\sim$150 K \citep{rof11} and the gas density was 7$\times$ 10$^{7}$ cm$^{-3}$ \citep{rof11}. Therefore, the two-phase warm-up chemical model of \citet{suz16}, which is based on the time scale, is appropriate for explaining the chemical abundance and evolution of CH$_{2}$NH towards G10.47+0.03. After the simulation, \citet{suz16} observed that the modelled abundance of CH$_{2}$NH varied between $\sim$10$^{-9}$--10$^{-8}$ in the gas phase. Similarly, the abundance of CH$_{2}$NH on the grain surface is $\leq$10$^{-12}$. We found that the abundance of CH$_{2}$NH towards the G10.47+0.03 is 2.61$\times$10$^{-8}$, which is nearly similar to the modelled abundance of CH$_{2}$NH in the gas phase derived by \citet{suz16}. This result indicates that CH$_{2}$NH was created towards G10.47+0.03, via the gas-phase neutral-neutral reaction between \ce{CH3} and NH radicals.

\begin{figure*}
	\centering
	\includegraphics[width=0.9\textwidth]{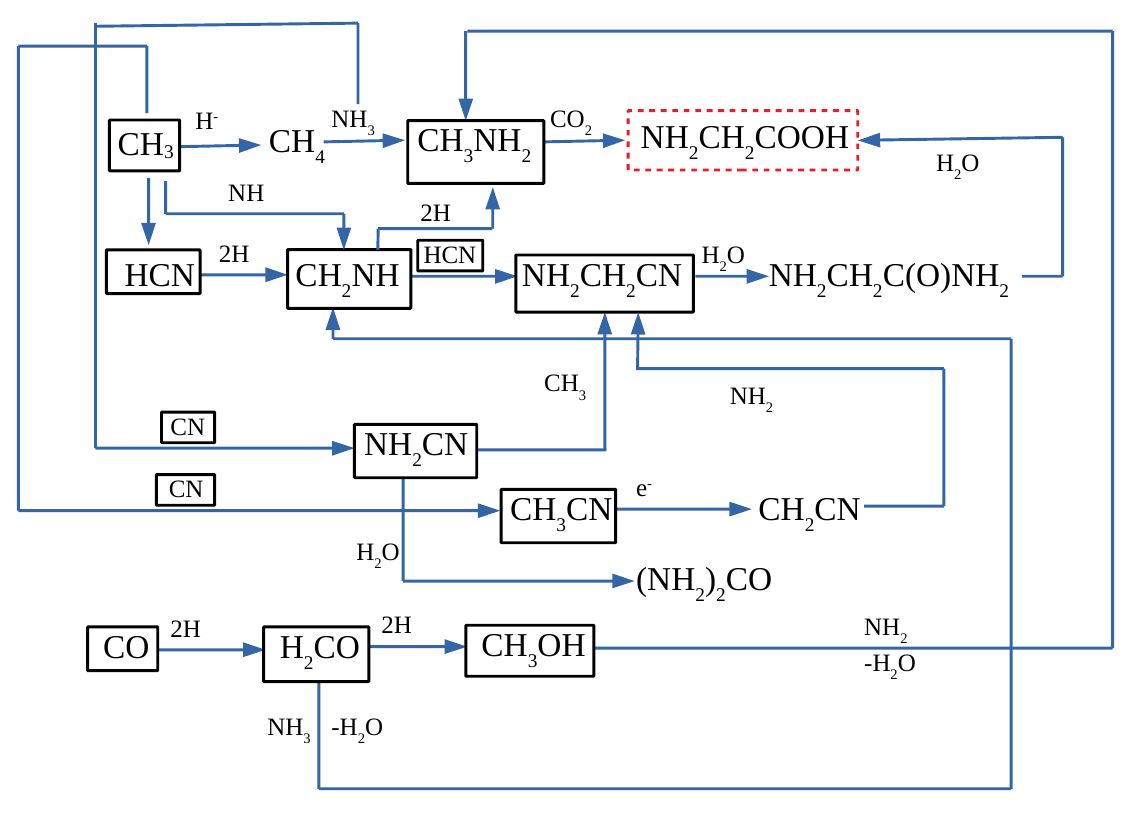}
	\caption{Proposed chemical network for the formation of \ce{NH2CH2COOH} from other molecules. In the network, the red box molecule is the final daughter molecule (\ce{NH2CH2COOH}) and the black box molecules are the parent molecules that are detected towards G10.47+0.03.}
	\label{fig:network}
\end{figure*}

Previously, \citet{man22a} claimed that \ce{NH2CH2CN} was the daughter molecule of \ce{CH2NH} (see Figure~\ref{fig:reaction}). The identification of both \ce{CH2NH} and \ce{NH2CH2CN} indicates that G10.47+0.03 is an ideal candidate in the ISM, where \ce{NH2CH2COOH} may exist. In ISM, G10.47+0.03 is the only source where the maximum number of possible \ce{NH2CH2COOH} precursors (such as \ce{NH2CN}, \ce{H2CO}, \ce{CH3NH2}, \ce{CH2NH}, and \ce{NH2CH2CN}) is detected, and several prebiotic chemistries have been proposed to understand the possible formation mechanism of these molecules and their possible connection with \ce{NH2CH2COOH}. After detecting the maximum number of \ce{NH2CH2COOH} precursors towards G10.47+0.03, we created a possible chemical network to understand the prebiotic chemistry of \ce{NH2CH2COOH} towards G10.47+0.03. The chemical network is shown in Figure~\ref{fig:network}. In the chemical network, all reactions were obtained from \cite{wo02, thu11, dan11, gar13, alo18, oh19, man22a} and UMIST 2012 astrochemistry molecular reaction databases. The chemical network clearly indicates the maximum number of parent molecules detected towards G10.47+0.03, which gives us an idea about the chemical complexity towards hot molecular cores.

\subsection{4.3. Searching of \ce{NH2CH2COOH} towards the G10.47+0.03 using the ACA}
After the identification of three possible \ce{NH2CH2COOH} precursor molecules like \ce{CH2NH} (present paper), \ce{CH3NH2} \citep{oh19}, and \ce{NH2CH2CN} \citep{man22a} towards the G10.47+0.03, we searched the emission lines of
\ce{NH2CH2COOH} conformers I and II towards the G10.47+0.03. After the careful spectral analysis using the LTE model, we did not detect any evidence of \ce{NH2CH2COOH} conformers I and II towards the G10.47+0.03 within the limits of our LTE modelling. The estimated upper limit column density of \ce{NH2CH2COOH} conformers I and II towards G10.47+0.03 was $\leq$1.02$\times$10$^{15}$ cm$^{-2}$ and $\leq$2.36$\times$10$^{13}$ cm$^{-2}$ respectively. The energy of \ce{NH2CH2COOH} conformer I is 705 cm$^{-1}$ (1012 K) lower than that of \ce{NH2CH2COOH} conformer II \citep{lov95}. The dipole moments of \ce{NH2CH2COOH} conformer I are $\mu_{a}$ = 0.911 D (a-type) and $\mu_{b}$ = 0.607 D (b-type), whereas \ce{NH2CH2COOH} conformer II has dipole moments of $\mu_{a}$ = 5.372 D (a-type) and $\mu_{b}$ = 0.93 D (b-type) \citep{lov95}. In ISM, the detection of a-type\footnote{The a-type and b-type are the different transitions of \ce{NH2CH2COOH}, whose spectral parameters depend on the different electric dipole moments.} transitions of \ce{NH2CH2COOH} are expected compared to b-type transitions because the line intensity of the molecule is proportional to the square of the dipole moments \citep{lov95}. The detection of three possible \ce{NH2CH2COOH} precursor molecules towards G10.47+0.03 gives more confidence about the presence of \ce{NH2CH2COOH} towards G10.47+0.03.

\section{5. Conclusion}
\label{conclu}
In this article, we present the identification of the possible \ce{NH2CH2COOH} precursor molecule \ce{CH2NH} towards G10.47+0.03, using the ACA band 4. The main conclusions of this study are as follows: \\\\
1. We successfully identified three non-blended transition lines of \ce{CH2NH} towards the G10.47+0.03 using the ACA observation. \\\\
2. The estimated column density of \ce{CH2NH} towards G10.47+0.03 is (3.40$\pm$0.2)$\times$10$^{15}$ cm$^{-2}$ with a rotational temperature of 218.70$\pm$20 K. The estimated fractional abundance of \ce{CH2NH} towards the G10.47+0.03 with respect to \ce{H2} is 2.61$\times$10$^{-8}$.\\\\
3. We compare the estimated abundance of \ce{CH2NH} with the two-phase warm-up chemical model abundance of \ce{CH2NH} proposed by \citet{suz16}. We noticed that the modelled abundance of \ce{CH2NH} is nearly similar to the observed abundance of \ce{CH2NH} towards G10.47+0.03. This comparison indicates that \ce{CH2NH} is created via a gas-phase neutral-neutral reaction between \ce{CH3} and NH radicals towards G10.47+0.03. \\\\
4. After the successful detection of \ce{CH2NH} towards G10.47+0.03, we also search the emission lines of the simplest amino acid \ce{NH2CH2COOH} conformers I and II towards G10.47+0.03. We do not detect the emission lines of \ce{NH2CH2COOH} conformers I and II within the limits of LTE modelling. The estimated upper-limit column densities of \ce{NH2CH2COOH} conformers I and II are $\leq$1.02$\times$10$^{15}$ cm$^{-2}$ and $\leq$2.36$\times$10$^{13}$ cm$^{-2}$ respectively. \\\\
5. The unsuccessful detection of \ce{NH2CH2COOH} towards G10.47+0.03 using ACA indicate that the emission lines of \ce{NH2CH2COOH} may be below the confusion limit in G10.47+0.03.

\ack[Acknowledgement]{We thank the anonymous referees for their helpful comments, which improved the manuscript. A.M. acknowledges the Swami Vivekananda merit cum means scholarship (SVMCM), Government of West Bengal, India, for financial support for this research. The plots within this paper and other findings of this study are available from the corresponding author upon reasonable request. This paper makes use of the following ALMA data: ADS /JAO.ALMA\#2016.2.00005.S. ALMA is a partnership of ESO (representing its member states), NSF (USA), and NINS (Japan), together with NRC (Canada), MOST and ASIAA (Taiwan), and KASI (Republic of Korea), in co-operation with the Republic of Chile. The Joint ALMA Observatory is operated by ESO, AUI/NRAO, and NAOJ.}

\section*{Conflicts of interest}The authors declare no conflict of interest.


\end{document}